\documentclass[twocolumn]{article}

\usepackage[modulo,switch]{lineno}
\modulolinenumbers[1]

\usepackage[latin9]{inputenc}
\usepackage{amsmath}
\usepackage{amssymb}
\usepackage{graphicx}
\usepackage{subfig}
\usepackage{wrapfig} 

\makeatletter\@ifundefined{date}{}{\date{}}
\makeatother

\evensidemargin\oddsidemargin \hoffset-22.4mm \voffset-28.4mm
\begin{document}

\title{Three-dimensional simulations of sound propagation in a trumpet with accurate mouthpiece shank geometry}

\author{Janelle Resch$^{1)}$, Lilia Krivodonova$^{2)}$, John Vanderkooy$^{3)}$ \\
$^{1, 2)}$ Applied Mathematics, University of Waterloo, Ontario, Canada N2L 3G1.\\ jresch@uwaterloo.ca, lgk@uwaterloo.ca\\
$^{3)}$ Physics and Astronomy, University of Waterloo, Ontario, Canada N2L 3G1.\\
jv@uwaterloo.ca.}

\maketitle\thispagestyle{empty}

\begin{abstract}
The length and bore geometry of musical instruments directly influences the quality of sound that can be produced. In brass instruments, nonlinear effects from finite-amplitude wave propagation can lead to wave distortion giving sounds a brassy timbre \cite{FMA, Blackstock, SWTrom, Myers, Rendon}. In this paper, we propose a three-dimensional model to describe nonlinear wave propagation in a trumpet and investigate the importance of the mouthpiece shank geometry. Time pressure waveforms corresponding to $B_3^b$ and $B_4^b$ notes were recorded at the mouthpiece shank and used as inputs for our model. To describe the motion of compressible inviscid fluid, we numerically solved the compressible Euler equations using the discontinuous Galerkin method. To validate our approach, the numerical results were compared to the recorded musical notes outside the bell of the trumpet. Simulations were performed on computational trumpets where different bore geometries were considered. Our results demonstrate that the shape of the narrow region near mouthpiece greatly influences the wave propagation and accuracy of the trumpet model.
\end{abstract}

\vspace{-3mm}

\section{Introduction}

\vspace{-1mm}

Finite-amplitude sound waves play a significant role in understanding the acoustics of brass instruments, and various models have been proposed to describe the corresponding wave propagation \cite{Blackstock}. Often in the literature, the tubing before the bell of the trumpet (and trombone) is assumed to be cylindrical with a constant radius, especially if the mathematical description of sound wave propagation is reduced to two dimensions or even one dimension. This assumption is a natural starting point to model a brass instrument since finite-amplitude standing waves and acoustic wave propagation in uniform cylindrical tubes have been studied for quite some time. Some of the initial work was done by Weiner \cite{W} in 1966. He examined the consequences of nonlinear wave propagation in air columns such as wave steepening and the formation of shock waves. Further discussions can be found in \cite{Non} and \cite{Non2}. 

This initial work on finite-amplitude wave propagation in tubes initiated some of the more recent approaches to describe nonlinear wave propagation in brass instruments. For instance, some models have been proposed where the input impedance or radiation impedance of the instrument are prescribed as boundary conditions. In theory, this approach seems quite reasonable since the peaks of the impedance curves correspond to the resonant frequencies of the bore, which characterizes acoustic behaviour \cite{ws}. However, the impedance either has to be measured or calculated by using acoustic pulse reflectometry methods which traditionally assumes that the cylindrical bore leading up to the flare is uniform \cite{KS}. This means that impedance formulations can be problematic for brass instruments (not as much for woodwind instruments) because the slope of the horn expansion can quickly become too large as discussed in \cite{FDTD} and \cite{waveprog}. In \cite{waveprog} however, Bilbao \textit{et al.} examined an alternative description of the radiation impedance that was presented in earlier findings \cite{radimp} to model wave propagation and radiation in a trumpet and trombone. The radiation impedance from \cite{radimp} takes the bore's curvature into account. Bilbao \textit{et al.} consider a finite difference scheme to compare a one-dimensional (1D) plane and spherical model using a transmission-matrix method. Though it was not explicitly stated if the bore near the mouthpiece end was assumed to be constant or not, their model was applied to linear wave propagation.

Frequency-domain models have also been a common approach to describe wave propagation in musical instruments. For example, Gilbert \textit{et al.}, considered a frequency -domain model based on the generalized Burgers' equation \cite{STBS}. Simulations were performed on geometries constructed after a Courtois bass trombone where the bore before the flare was approximated by a cylindrical tube. Another recent model discussed by Noreland \textit{et al.} in \cite{hybrid} introduces a hybrid scheme to model wave propagation in brass musical instruments. For the flare region, a two-dimensional (2D) finite-element method was considered where the inviscid Helmholtz equation was used. The cylindrical part of the instrument before the flare was analyzed using a 1D viscid transmission line model. This bore section was further separated into two parts (but modelled in the same way). The first section was near the mouthpiece boundary and the radius was assumed to be constant until a slow flare begins in the second segment. Although the model seemed reasonable for low frequency notes, the authors concluded that for high frequencies, a more accurate model was needed. 

A frequency model that does consider some of the geometric features near the mouthpiece of a brass instrument is proposed by Thompson \textit{et al.} in \cite{IWSFDMTSP}. The authors considered a linear and nonlinear frequency-domain model for the trombone where the shape was approximated by a sequence of 152 cylindrical tubes. In particular, some radius variation were considered in the initial 7.85 cm of tubing. For the next 53.14 cm of tubing before the flare however, the radius of the cylinder is assumed to be constant. It was found that for the loudly played notes, the nonlinear model matched the fundamental frequency and second harmonic of the experimental data, and followed the general shape of the spectral curve for components greater than 8000 Hz. However, the model deviated for frequency components in between. 

Previously in \cite{Ourpaper}, we too assumed that the bore before the trumpet bell was uniform and modelled the wave propagation of measured musical notes by numerically solving the full 2D compressible Euler system using the discontinuous Galerkin (DG) method. The consequences of neglecting the spatial dimension was also examined. Our simulations produced a similar harmonic distribution to the notes for components that were transmitted from the bell. However, the amplitude was overestimated and there were discrepancies in the lower spectra. We then considered the analogous model in three dimensions and presented the initial findings in \cite{Ourpaper2}. Modelling the sound propagation with both spatial dimensions greatly improved our results, however, the resulting amplitude of the simulated notes was approximately 8 dB too high. However, close inspection of the bore near the mouthpiece reveals that the inner tubing before the first bend is more complicated than alluded to by the outer geometric features of the instrument, i.e., we found that the tube of the trumpet near the mouthpiece was not cylindrical. In fact, the radius of the trumpet bore from the mouthpiece varies for approximately 24 cm. 

Therefore, in this paper, we propose a three-dimensional (3D) model for nonlinear sound wave propagation in a trumpet where we attempt to take the variations of the narrow bore near the mouthpiece into account. It is our belief that these subtle geometric features need to be represented to accurately construct and simulate the evolution of nonlinear waves inside the instrument, and perhaps, may be more significant than energy losses due to viscous friction or wall vibrations. As Rend\'{o}n states,  the instrument ``determines the extent to which the bore profile is able to support nonlinear propagation'' \cite{Rendon}.  For any region in the instrument in which the cross-sectional area increases, the amplitude of the propagating waves will decrease because the sound energy spreads and the particle velocity decreases \cite{Myers}. These aspects will in turn influence the nonlinear standing wave pattern. Therefore, in our purposed model, we attempted to represent the initial bore and incorporate nonlinearities as well as compressibility effects. To achieve this, the compressible Euler equations found in gasdyanmics were used where viscosity and other losses were neglected. The equations of motion were then solved numerically using the DG method, and the corresponding numerical results will be shown and discussed in Section \ref{s:res}.



\vspace{-3mm}

\section{Model}

\vspace{-1mm}

\noindent \subsection{{Equations of Motion}}
\vspace{-1mm}

The 3D compressible Euler equations are defined as:

\begin{equation}\label{eq:euler}
\begin{bmatrix}\rho \\ \rho u \\ \rho v \\ \rho  w \\ E \end{bmatrix} _t+\begin{bmatrix}\rho u \\ \rho u^2 + p \\  \rho u v \\ \rho u w \\ u(E+p)   \end{bmatrix} _x +  \begin{bmatrix}  \rho v \\ \rho u v \\ \rho v^2 +p \\ \rho v w \\v(E+p) \end{bmatrix} _y+   \begin{bmatrix}  \rho w \\ \rho u w \\ \rho v w \\ \rho w^2 + p\\w(E+p) \end{bmatrix}_z  =0,
\end{equation}

\noindent where $\rho$ is the gas density; $(\rho u, \rho v, \rho w)$ are the momenta in the x, y and z direction, respectively; p is the internal pressure; and E is the total energy. Finally, the parameter $\gamma$ is the specific heat ratio which is $\gamma \approx 1.4$ for air \cite{NDGM}. 

The equation of state for an ideal gas connects E to the other variables and closes the system
\begin{equation}\label{eq:state}
E=\frac{p}{\gamma - 1} +\frac{\rho}{2}(u^2+v^2+w^2).
\end{equation}
\noindent The system (\ref{eq:euler}) and (\ref{eq:state}) describes the motion of inviscid, compressible gas. We used this mathematical model to simulate musical notes which were played by a live musician. 


\noindent \subsection{Setup and Experimental Data} \label{s:exp}

\begin{figure}  
\centering
\includegraphics[scale=1.1, width=7.5cm]{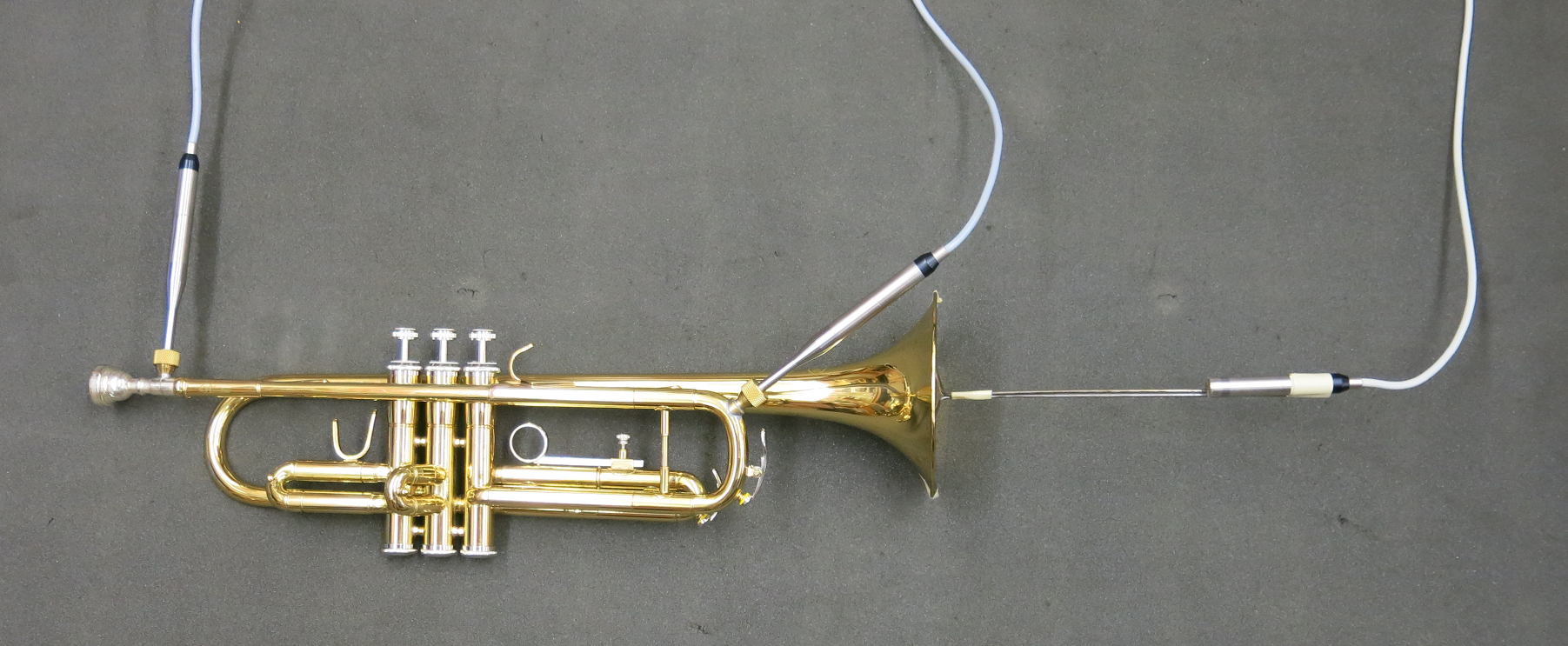}
\caption{(Colour online) Placement of microphones on the Barcelona BTR-200LQ trumpet.} \label{experimentphoto2}
\label{fi:mouthpiece}
\end{figure}

\begin{figure} [!ht]
\centering
\subfloat{\includegraphics[scale=0.5, width=7.5cm]{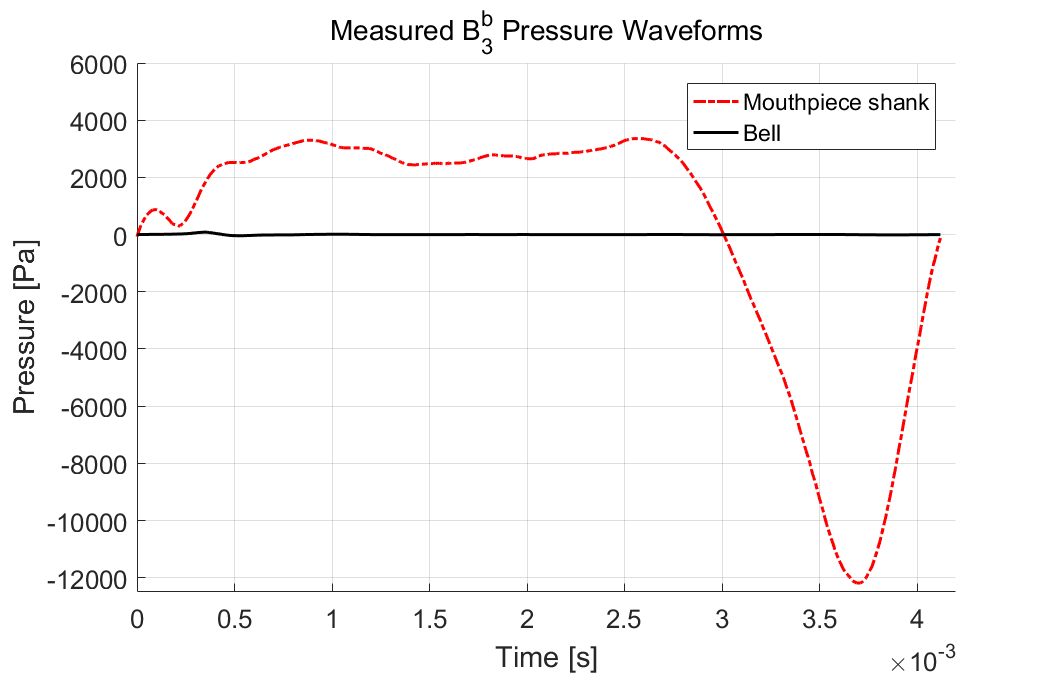}}
\qquad
\subfloat{\includegraphics[scale=0.5, width=7.5cm]{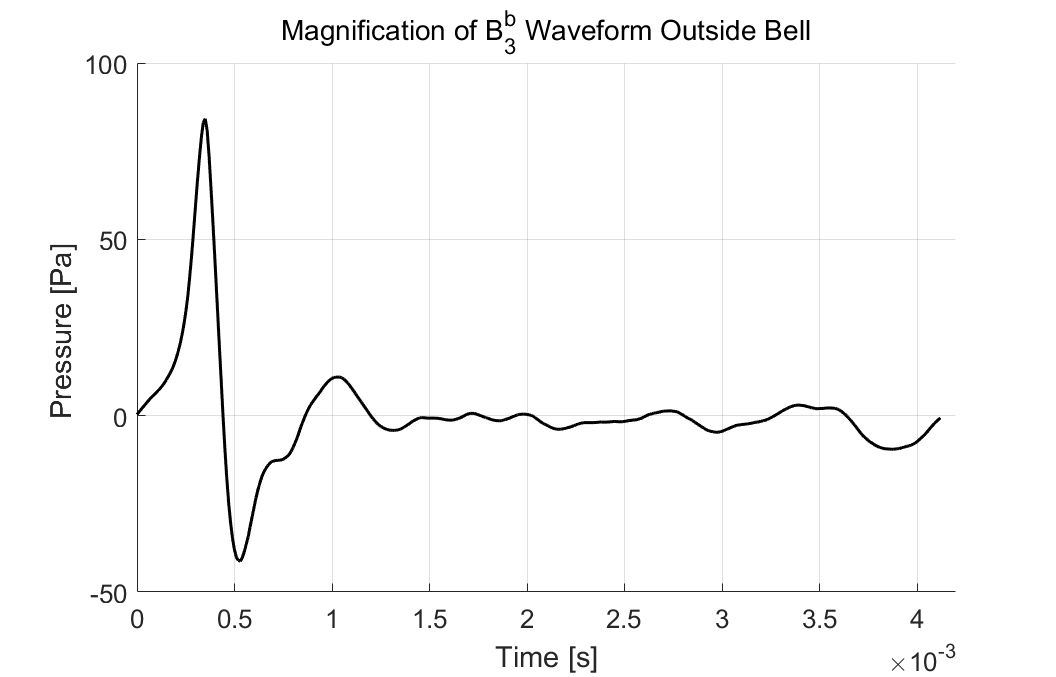}} 
\caption{(Colour online) Top: Measured pressure waveform of the $B_3^b$ at the mouthpiece shank and outside the bell. Bottom: Magnification of the measurements outside the bell.}
\label{fi:B3}
\end{figure}

\begin{figure}[!ht]
\centering
\subfloat{\includegraphics[scale=0.5, width=7.5cm]{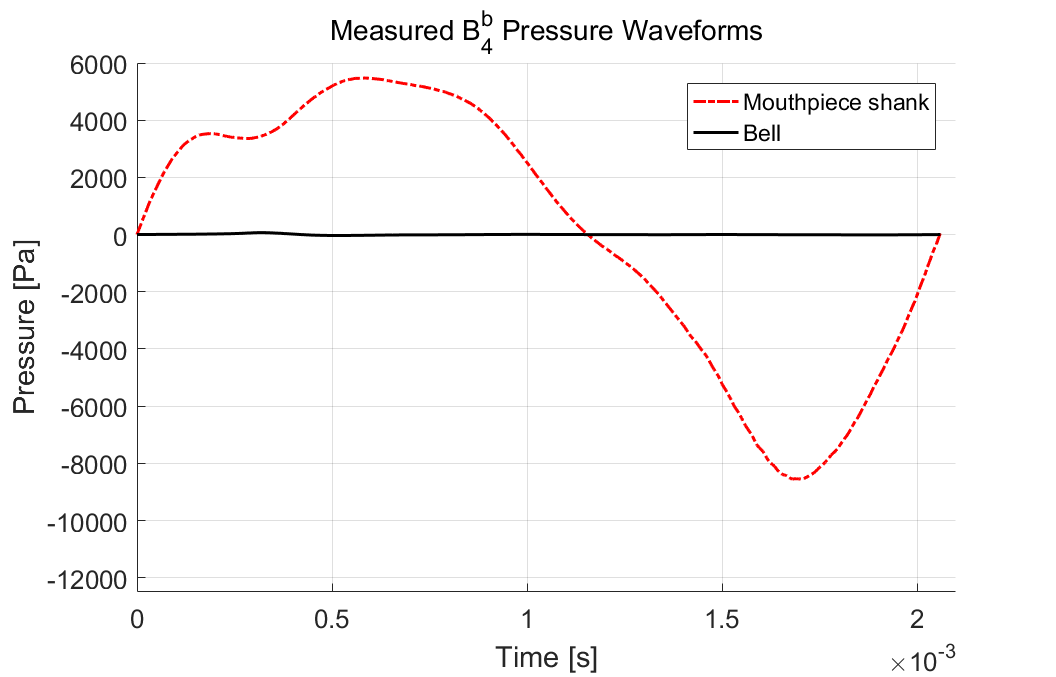}}
\qquad
\subfloat{\includegraphics[scale=0.5, width=7.5cm]{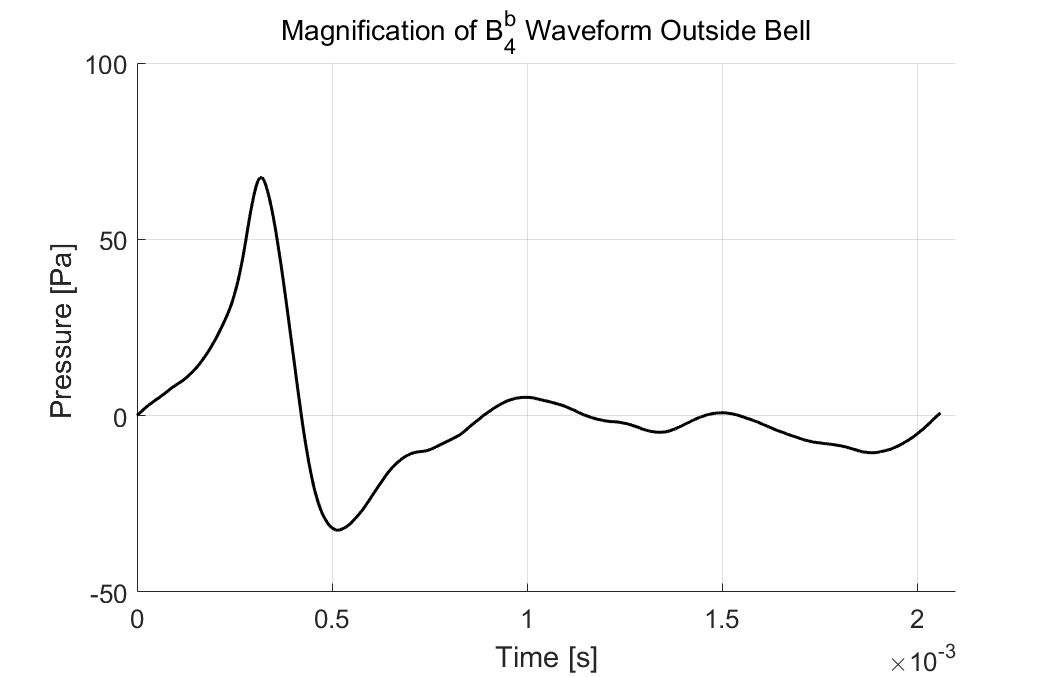}}
\caption{(Colour online) Top: Measured pressure waveform of the $B_3^4$ at the mouthpiece shank and outside the bell. Bottom: Magnification of the measurements outside the bell.}
\label{fi:B4}
\end{figure}

Sound pressure measurements of the $B_3^b$ and $B_4^b$ notes played at \textit{forte} on the $B^b$ Barcelona BTR-200LQ trumpet shown in Figure \ref{experimentphoto2} were collected at two locations. A quarter-inch microphone was mounted near the shank of the mouthpiece, and a half-inch microphone was placed on the central axis of the trumpet roughly 16 cm outside the bell. For convenience, we will refer to these locations as \textit{mouthpiece} and \textit{bell}. The top plots of Figures \ref{fi:B3} and \ref{fi:B4} depict one period of the recorded time pressure waveforms for the $B_3^b$ and $B_4^b$ notes, respectively. Since the pressure at the bell is much lower than inside the instrument, the bottom plots of Figures \ref{fi:B3} and \ref{fi:B4} show a magnification of the bell measurements. All pressure plots show the deviation from atmospheric pressure. 

\begin{figure} [!ht]
\centering
\subfloat{\includegraphics[scale=0.5, width=7.5cm]{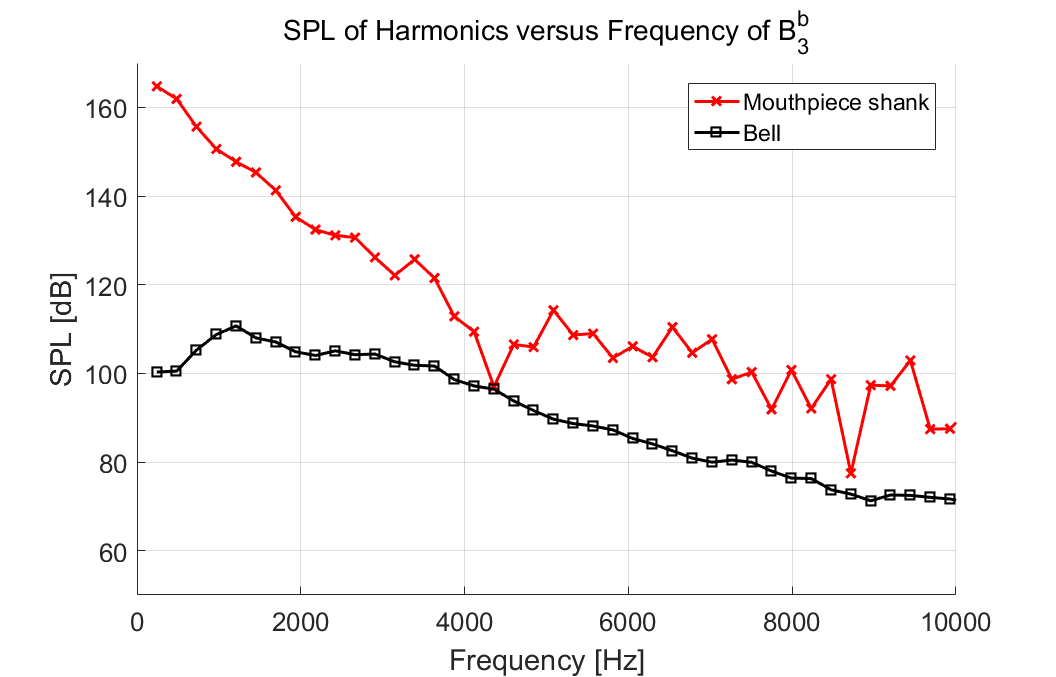}}
\qquad
\subfloat{\includegraphics[scale=0.5, width=7.5cm]{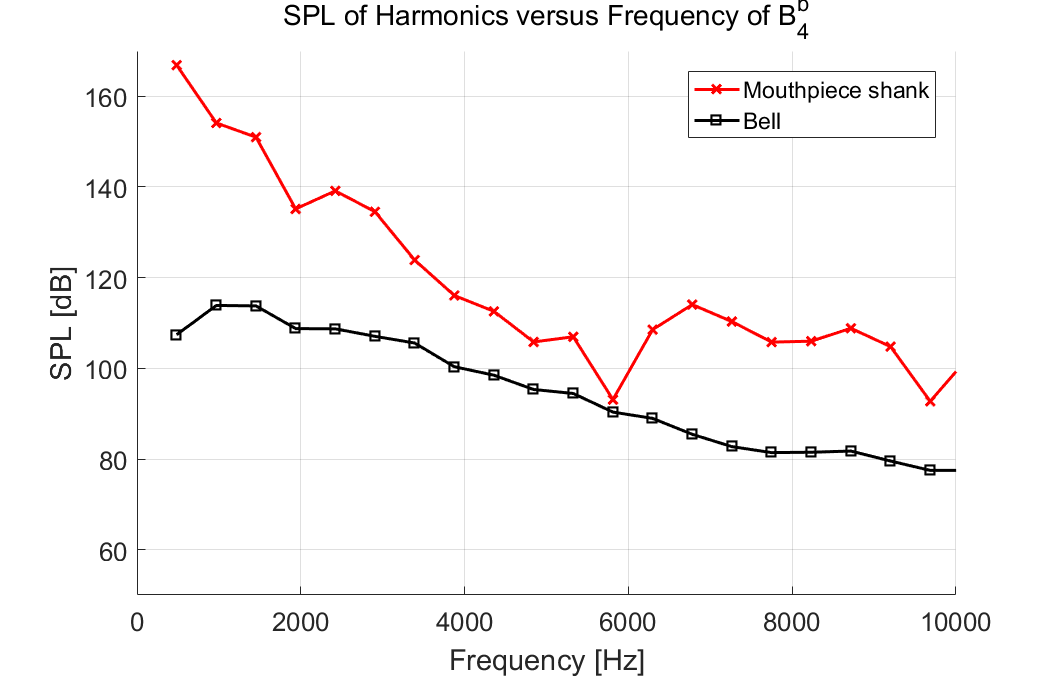}}
\caption[]{(Colour online) Frequency spectra at the mouthpiece shank and outside the bell. Top: $B_3^b$. Bottom: $B_4^b$. }
\label{fi:B34FFT}
\end{figure}

The experimental data collected at the mouthpiece will be used as the boundary condition on pressure for numerical simulations. The data at the bell will then be used to judge the accuracy of our proposed model. As customary, the results are presented as sound pressure level (SPL) which is measured in decibels (dB). The total sound pressure level in the presented examples is equal to 166.9 dB for the $B_3^b$ and 167.2 dB for the $B_4^b$ at the mouthpiece position. Dynamic levels in musical notation, i.e., \textit{piano}, \textit{forte}, etc., are arbitrary in the sense that there is no specific decibel level that defines \textit{forte}. But typically, the SPLs we obtained from our experiments fall into the \textit{forte} range reported in the literature. Our obtained results are similar in character to the pressure measurements presented in \cite{HGT, RBWI, SWTrom, Norman, IWSFDMTSP}. In particular, the waveform shapes and SPLs for analogous notes in these papers resemble our plots shown in Figures \ref{fi:B3}, \ref{fi:B4} and \ref{fi:B34FFT}.

The signals from the microphones mounted on the trumpet were recorded on a four-channel Tektronix oscilloscope.  Its resolution was 8 bits, and the output sampling rate was 50 kHz.  The actual sampling rate of the oscilloscope is very high, and internal sample averaging gives an effective resolution of about 11 bits.  For bipolar signals this gives a signal-to-noise ratio of about 70 dB.  The spectra shown in our paper can have much more contrast than that because the noise is spread fairly uniformly over all frequency bins.  Each period analyzed by the discrete Fourier transform (DFT) has about 413 samples, so the noise level in each bin can be about 26 dB less than the straight 70 dB offered by the resolution of each sample.  We find that the acoustic noise in our measurements is always much less than the expected 96 dB contrast of the measurement system.

\begin{figure} [ht]
\centering
\includegraphics[width=7.5cm]{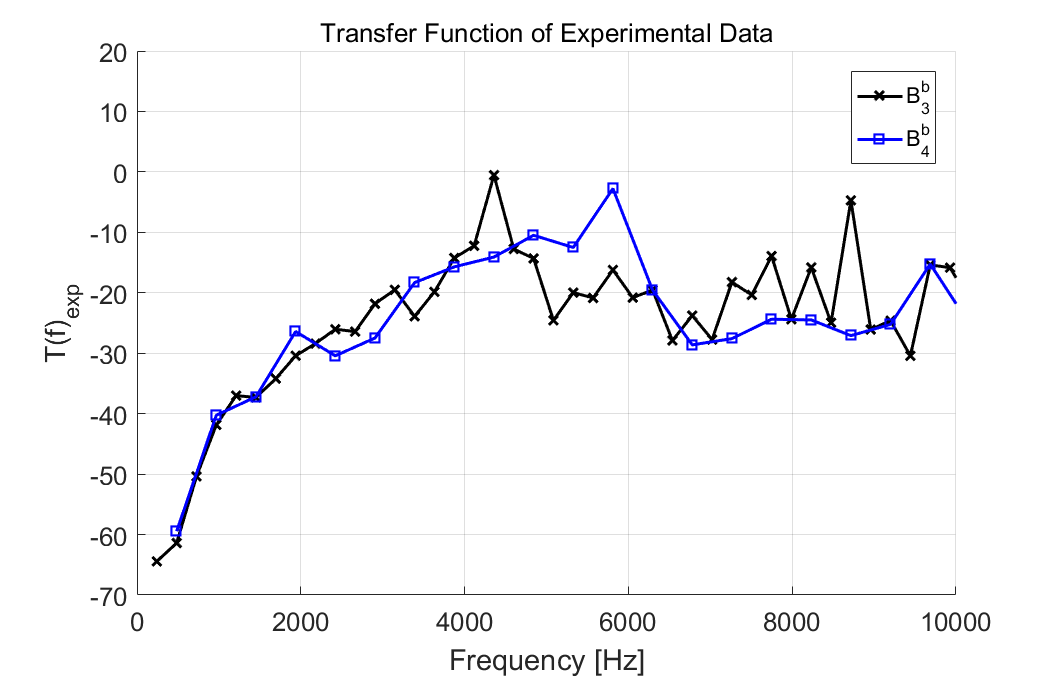} 
\caption[]{(Colour online) Transfer function of experimental data, specifically the ratio of the pressure measured outside the bell over the pressure measured at the mouthpiece.}
\label{fi:TF}
\end{figure}

The transfer function of the experimental data, denoted by $T(f)_\text{Exp}$, is a function of frequency (i.e., a frequency-domain measurement), denoted by f, and is defined as 

\begin{equation}\label{eq:transferFunction}
T(f)_\text{Exp}=\frac{P(f)_\text{bell}}{P(f)_\text{mouth}},
\end{equation}

\noindent where $P(f)_\text{bell}$ is the pressure measured outside the bell, and $P(f)_\text{mouth}$ is the acoustic pressure measured at the mouthpiece shank. In decibels, the transfer function is given by $20 \log_{10} \left (T(f)_\text{Exp} \right )$ \cite{Beau}. The corresponding function is plotted for the $B_3^b$ and $B_4^b$ in Figure \ref{fi:TF} and represents the power transmitted from the bell. The transfer function for both notes displays similar character for frequencies up to 4000 Hz. The remaining portion of the signal however almost appears to be noisy due to the observed zig-zag pattern, especially for harmonics greater than 6000 Hz. Therefore, from studying Figures \ref{fi:B34FFT} and \ref{fi:TF}, we are confident that the data up to 4000 Hz is reliable and does not contain much noise. 

Applying Fourier synthesis to the results in Figure \ref{fi:B34FFT}, we obtain a continuous expression for the pressure with respect to time. We write the trumpet notes as a sum of sinusoidal waves with respect to the fundamental frequency, denoted by $f$, and corresponding harmonics, $2f$, $3f$,$\dots$, each with a corresponding amplitude, denoted by A, and phase shift, denoted by $\phi$. Therefore, one period of the entire pressure waveform of a desired note is expressed as
\begin{equation}\label{eq:FFT}
p= A_0 + 2\sum_{i=1}^{N/2} A_i \text{cos} \left(2 \pi f_i t+\phi_i \right),
\end{equation}
\noindent where $A_0$ is the term corresponding to the direct current, N is the number of points in our data, and $f_{N/2}$ is the Nyquist frequency \cite{Gren}. For each measured note, the number of samples per period was rounded to 413. If an integer is not naturally obtained for the number of samples per period, rounding the value can influence the frequency spectrum. However, for our simulations, we were constructing the time pressure waveform with 30 harmonics, so we do not expect this rounding to make a difference for the corresponding spectral components.

\noindent \subsection{{Numerical Method}}

The equations of motion (\ref{eq:euler}) form a hyperbolic system of conservative laws which we solved numerically using the discontinous Galerkin (DG) method. The simulations were run on a graphics processing unit (GPU) using NVIDIA's Compute Unified Device Architecture (CUDA). Details about the algorithm for this implementation can be found in \ref{MLA}. The implementation takes advantage of the numerical method's parallelization features \cite{MLA}.

We used a second order accurate linear approximation in space and Runge-Kutta (RK2) discretization in time with the local Lax-Friedrichs Riemann solver. Using a higher order approximation does not seem necessary to improve the accuracy of our numerical results since the mesh resolution must be fine enough near the mouthpiece and bell to resolve these fine geometric features and capture the desired frequency components of the musical notes we will be simulating. Thus, improving the numerical accuracy over that of the experiment did not seem worthwhile. 

We also needed to ensure there were enough mesh elements in the radial direction to obtain accurate simulations. The meshes we used ranged from 603,201 to 1,317,219 elements. So along the horizontal direction, we have approximately 1980 tetrahedral elements per wavelength where the minimum radius is 0.0038392 mm. This was deemed sufficiently fine. The corresponding mesh size spacing was chosen so that our numerical results visually did not vary is a finer mesh was used. All the simulations that we will present in this paper were run through SHARCNET.

\noindent \subsection{{Initial and Boundary Conditions}}

We assume the flow initially is at rest and introduce the sound waves via the boundary condition. All variables are scaled from physical values to values more convenient for computation. In particular, the ambient speed of sound $c_0$, which is approximately 343 m/s in air, and atmospheric pressure $p_0$, are scaled to be equal to 1. Assuming that the flow is isentropic, i.e., $c_0 = \sqrt{\frac{\gamma p_0}{\rho_0}}$, the initial density $\rho _0$ should then be taken to be 1.4.  In summary 
\begin{equation}\label{eq:IC}
\begin{bmatrix} p_0, \rho _0, u_0, v_0, w_0, E_0\end{bmatrix} = \begin{bmatrix} 1.0, 1.4, 0, 0, 0, 2.5 \end{bmatrix}.
\end{equation}	
On the inner and outer walls of the trumpet, excluding the mouthpiece boundary, reflective boundary conditions (solid wall boundary conditions) were prescribed. A ghost state was specified so that the normal velocity was reflected with respect to the wall, i.e., taken with a change of sign. The density, pressure and tangential velocity were unchanged from the corresponding values inside the cell. At the left vertical boundary of the bore which corresponds to the mouthpiece boundary, the ghost state was specified to be the inflow condition which is described below. Finally, along the far-field boundary, pass-through boundary conditions were used in which the ghost state was prescribed to be the initial state defined in (\ref{eq:IC}). The computational domain was large enough to ensure that propagated waves achieved a sufficiently small amplitude for complete pass-through of waves without reflection which can adversely influence the solution. We experimentally determined the size of the domain so that reflections would not influence the solution, specifically the waveform exiting the bell.

When imposing boundary conditions at the mouthpiece, we related velocity to pressure at the mouthpiece through the expression derived from linear acoustic theory for planar waves
\begin{equation} \label{eq:vel}
p - p_0 = \rho_0 cu.
\end{equation}
The linearization is justified since the speed of the air particles inside the instrument is low relative to the speed of sound. Velocity measurements for trombones reported in \cite{PMI} give the maximum speed about 17 m/s, i.e., about 5\% of the speed of sound, in the throat of the mouthpiece which are similar values observed in our numerical simulations. 

Finally, the density is computed assuming we have an adiabatic process. Thus, compressible flow theory states that
\begin{equation}
\rho = C p^{\frac{1}{\gamma}},
\end{equation}
\noindent where $C=\gamma$ is the proportionality constant \cite{EG}.

The experimental data obtained at the mouthpiece of the trumpet was written as a sum of $N_f=30$ sinusoidal waves and used as the boundary condition on pressure. We do not attempt to model the flow behaviour in the mouthpiece cup as it is quite complex. 
In summary, the dimensionless boundary conditions at the mouthpiece of the computational trumpet are given by 
\begin{equation}\label{eq:BC}
\text{} 
\begin{cases}
\hat{p}= \hat{A}_0 + \sum_{i=1}^{N_f} 2\hat{A}_i \text{cos} \left(2 \pi \hat{f}_i t+\hat{\phi}_i \right),\\
\hat{\rho} = \gamma \hat{p}^{\frac{1}{\gamma}},\\
\hat{u}=\frac{\hat{p}-p_o}{\rho_o c},\\
\hat{v}=0.0, \\
\hat{w}=0.0, \\
\hat{E} = \frac{\hat{p}}{\gamma -1} + \frac{\hat{\rho}}{2}(\hat{u}^2+\hat{v}^2+\hat{w}^2),
\end{cases}
\end{equation}
\noindent where $\hat{A}_i$, $\hat{f}_i$, and $\hat{\phi}_i$ denote the amplitude, frequency and phase shift, respectively, for each harmonic of the measured notes \cite{Ourpaper}. 

\noindent \subsection{Geometrical Trumpet Model} \label{meshSec}

We constructed three geometries to represent the 1.48 m long trumpet depicted in Figure \ref{experimentphoto2}. The focus was to accurately model the length of the trumpet, the bore, and slowly increasing diameter of the bell. In \cite{Ourpaper}, we justified that the bends of the instrument do not greatly influence the wave propagation, especially in comparison to the bell. Therefore, for all the numerical simulations discussed here, the bends will not be modeled, i.e., the bends and coils of the trumpet will be unwrapped and the tubing will be straightened out. 

The bell was modelled with great care since propagating sound pressure waves are especially sensitive to the bell's curvature because the expansion influences harmonic reflections. More precisely, the location at which the harmonic waves reflect in the bell is dependent on their frequency. As the frequency of the waves increase, a larger portion of the energy will be transmitted from the bell \cite{FMA}. This implies that slight inaccuracies in the bell geometry could produce exaggerated discrepancies in numerical simulations as we showed in \cite{Ourpaper2}. To obtain a realistic flare shape, a picture of the trumpet bell was taken. The \textit{grabit} software from Math Works Inc. was then used to trace out the trumpet flare by a series of points. 

We initially constructed a computational trumpet, which for convenience will be referred to as \textit{Geometry 1}, to have an uniform cylindrical bore prior to the flare. However, we questioned the validity of this simplification since in reality, the tubing near the mouthpiece is more complicated. From an observer's perspective, the radius of the bore from the mouthpiece appears to slowly increase for approximately 24 cm. Then, the tube seems to remain cylindrical until it begins to widen again 102 cm from the mouthpiece. The geometry of the mouthpiece itself is also quite complex and varies in shape and dynamics for each model. A general diagram of a brass instrument mouthpiece is shown in Figure \ref{fi:mouthpiece}. We tried to avoid some of these potential intricate effects by measuring the sound pressure waveforms at the shank of the mouthpiece. The corresponding microphone position is also depicted in Figure \ref{fi:mouthpiece}.

The mouthpiece throat radius is approximately 0.4$\times$ the shank radius; and the radius of the trumpet bore 20 cm from the shank is approximately 1.42$\times$ the shank radius. This change in volume seems rather significant and not considering these geometric attributes may cause an overestimation in the simulated amplitude. Furthermore, for such sections in the trumpet where the bore is expanding in diameter, the pressure amplitude of the forward moving waves decreases thereby reducing the cumulative nonlinear distortion \cite{france}. Moreover, just as the rapid flare of the trumpet influences harmonic reflections, we hypothesize that the increase in radius at the mouthpiece may have similar a consequence.

\begin{figure}  
\centering
\includegraphics[scale=1.1, width=7.5cm]{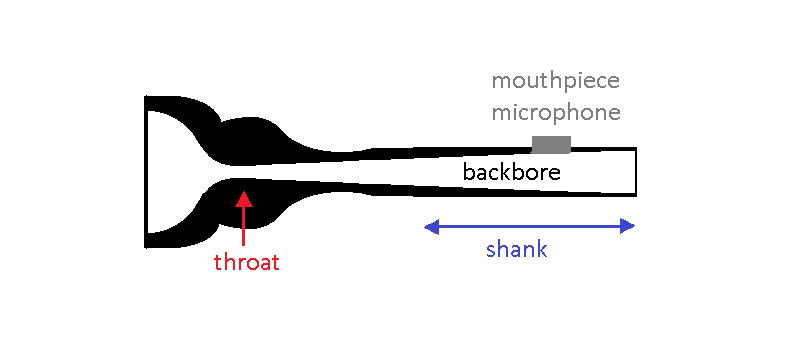}
\caption[]{(Colour online) A diagram of a trumpet mouthpiece.}

\label{fi:mouthpiece}
\end{figure}

\begin{figure} [ht]
\centering
\includegraphics[width=7.5cm]{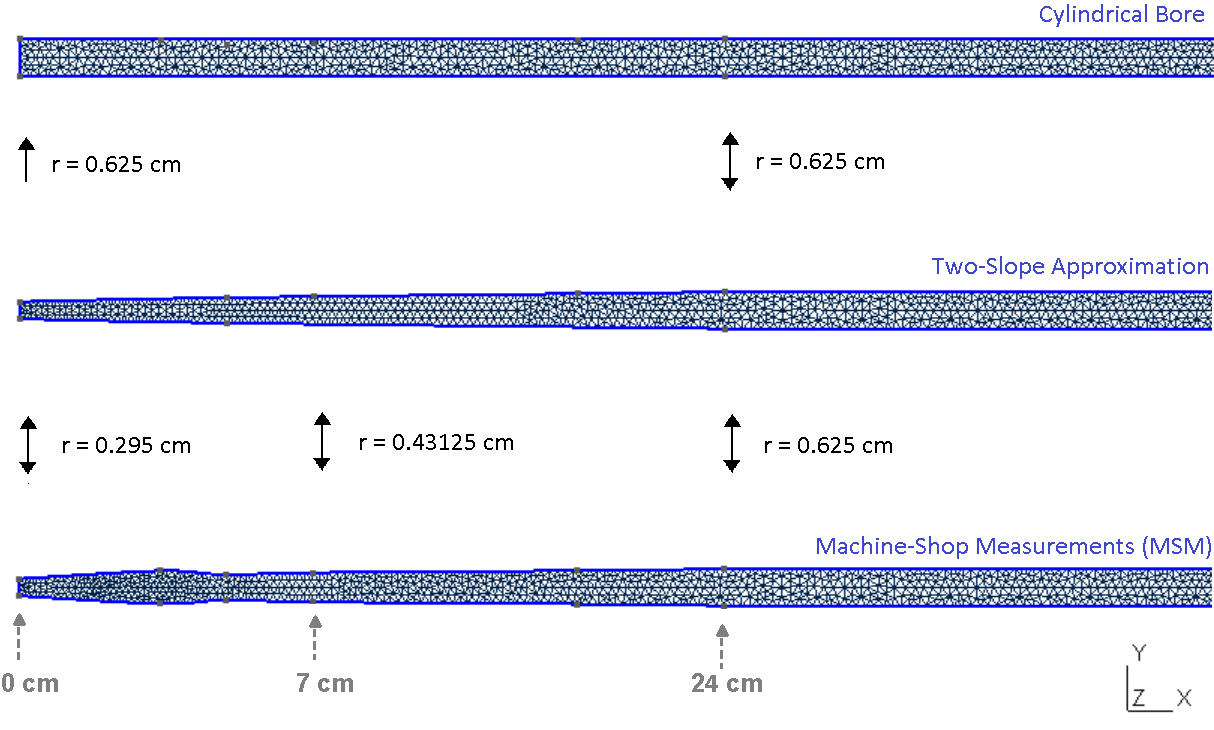} 
\caption[]{(Colour online) The geometric shapes of the tubing near from the mouthpiece boundary used to construct Geometries 1 to 3. Top: Corresponds to Geometry 1, with bore shape reference name ``cylindrical bore''. Middle: Corresponds to Geometry 2, with bore shape reference name ``two-slope approximation''. Bottom: Corresponds to Geometry 3, with bore shape reference name ``machine-shop measurements (MSM)''. Radii is the same at points indicated at the arrows.}
\label{fi:Bores}
\end{figure}

Therefore, we wanted to create a 3D representation of the trumpet where the variation found in the initial 24 cm of tubing is taken into account. We first measured the diameter of the tube in three positions: where the mouthpiece microphone was mounted (near the throat); at the end of the shank; and again where the tube radius appeared to stop changing. These three points were connected with cubic splines and we will refer to the corresponding trumpet geometry as \textit{Geometry 2}. While carrying out these measurements however, we observed that the change in diameter was much more complicated. Several measurements were then taken along the initial 24 cm of tubing at the university's machine shop. These points were also interpolated by cubic splines and the corresponding 3D representation of the trumpet will be called \textit{Geometry 3}. A cross-section of the different trumpet geometries can be seen in Figure \ref{fi:Bores}.

\noindent \subsubsection{{Summary of Trumpet Geometries}}

In summary, we constructed the following geometries with the following properties:
\begin{itemize}
\item[]Geo. 1: \textit{Cylindrical Bore}

The tubing from the mouthpiece boundary to beginning of bell's flare, i.e., $x \in [0 \text{ cm, } 102 \text{ cm}]$ is approximated by a cylindrical bore. 

\item[]Geo. 2: \textit{Two-Slope Approximation} 

The tubing from $x \in [0 \text{ cm, }  24\text{ cm} ]$ was approximated by three measurements: at the mouthpiece microphone position, the shank, and before the first bend. 

\item[]Geo. 3: \textit{Machine-Shop Measurements (MSM)} 

The tubing from $x \in [0 \text{ cm, }  24\text{ cm} ]$ was constructed from taking seven measurements.
\end{itemize}

For all geometries, the sets of points outlining the boundary of the computational trumpet were passed to the mesh generating software GMSH. The points on the straight region of the bore were connected by lines; the sets of points outlining the bell and the initial tubing for Geometries 2 and 3 were interpolated using cubic splines. The resulting curves were used to generate a 3D representation of the instrument via a rotational extrusion about $x=0$ to generate a spherically symmetric trumpet mesh. All meshes consist of tetrahedral elements with adaptive element sizes to accurately resolve the geometric features of the trumpet. The total number of cells for each mesh in addition to their properties can be found in Table \ref{ta:cells}.
\begin{center}
\begin{table}
\caption{Number of cells and properties associated with each trumpet mesh.}
\begin{tabular}{@{}| c | c | c |@{}}
\hline
Geometry &  Number of Cells & Bore Shape \\ \hline
Geometry 1  & 603,201 & Cylindrical  \\  
Geometry 2  & 1,302,915 & Two-Slope App.  \\  
Geometry 3  &  1,317,219 & MSM  \\
\hline
\end{tabular} \label{ta:cells}
\end{table}
\end{center}
\vspace{-3mm}

\noindent \section{{Simulation Results}} \label{s:res}
\vspace{-1mm}

\begin{figure}[!ht]
\subfloat{\includegraphics[width=7.5cm]{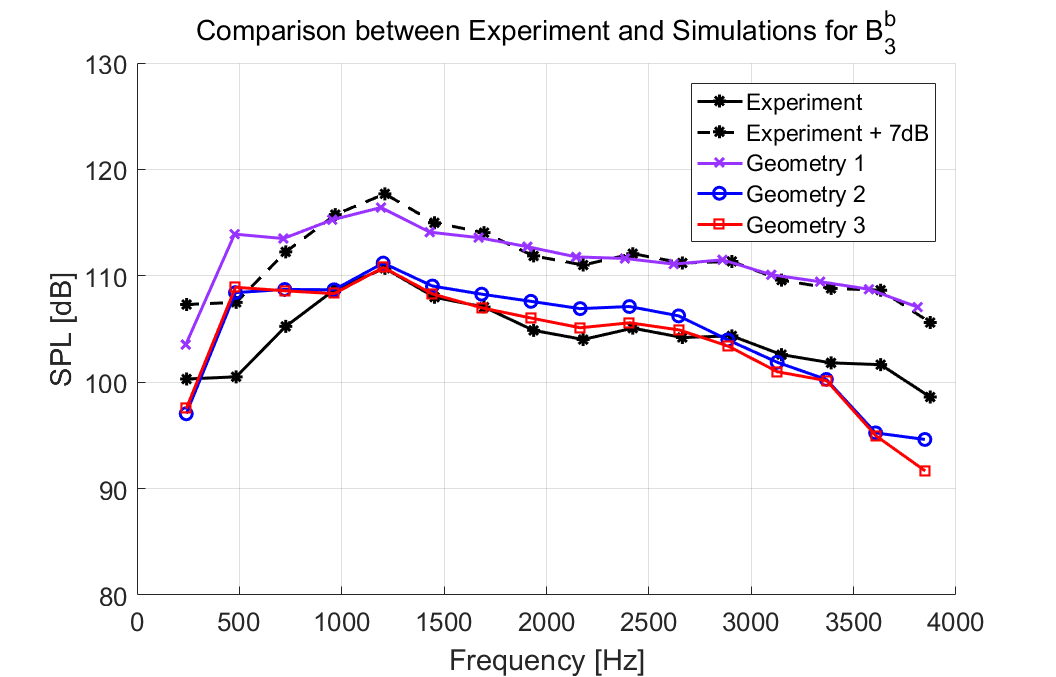}}
\qquad
\subfloat{\includegraphics[width=7.5cm]{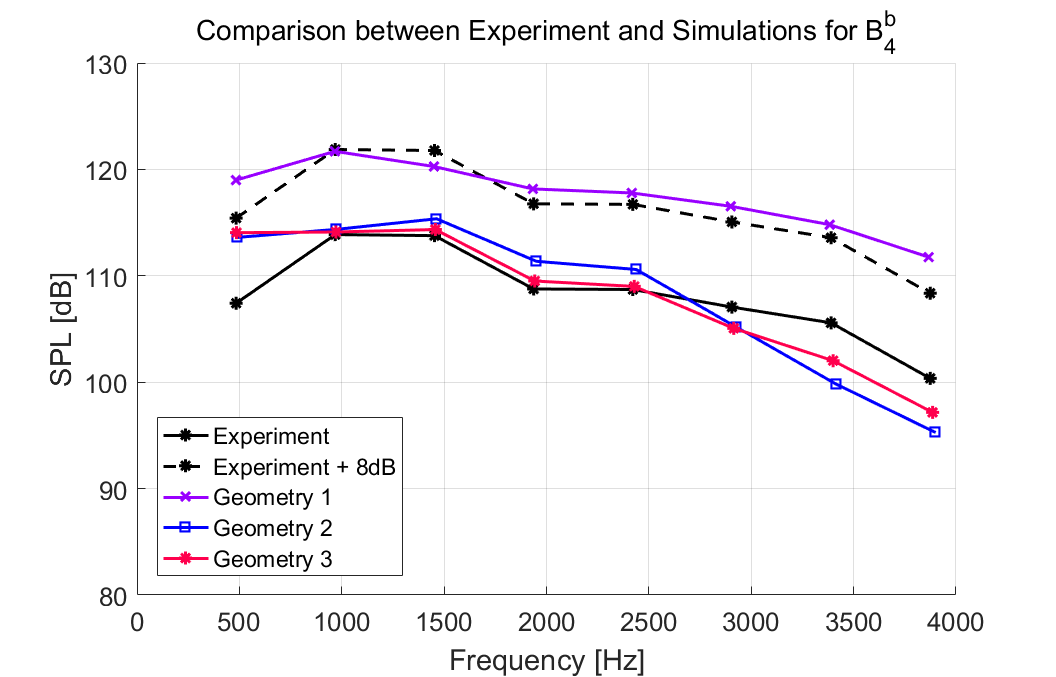}} 
\caption[]{(Colour online) Frequency spectra of experimental data and 3D simulation results for $B_3^b$ (top) and $B_4^b$ (bottom).}
\label{fi:3d}
\end{figure}

Simulations have been carried out to describe the nonlinear wave propagation of $B_3^b$ and $B_4^b$ notes (discussed in Section \ref{s:exp}) inside a trumpet. The 3D compressible Euler equations (\ref{eq:euler}) were solved with the initial and boundary conditions mentioned above on Geometries 1 to 3. The frequency spectra of the numerical output for the $B_3^b$ and $B_4^b$ notes sampled 16 cm outside the computational trumpet bell can be seen in the top and bottom plots of Figure \ref{fi:3d}, respectively. The frequency spectra of the experimental data at the bell position (which recall is 16 cm outside the bell) also is depicted and used to validate our model. 

For Geometry 1, a comparison of the numerical and experimental data shows that the numerical amplitude for the $B_3^b$ and $B_4^b$ notes is overestimated by approximately 7 dB and 8 dB, respectively. For comparative purposes, we shifted the experimental curves by the amplitude difference. When this shift is considered, we see that the harmonic distribution of the experimental data and numerical solutions are in good agreement for frequency components above 700 Hz.

For both notes, the resulting numerical amplitudes on Geometry 2 and Geometry 3 are now more closely representative of the experimental curves. We will first discuss the results obtained for the $B_3^b$ shown in the top plot of Figure \ref{fi:3d}. When Geometry 3 (which has the \textit{MSM} bore) is used, the numerical solution agrees with the experimental data almost perfectly for spectral components between 900 Hz - 3400 Hz. Similar results are obtained for Geometry 2 (which has the \textit{two-slope approximation} bore). However, compared to Geometry 3, the frequencies are slightly higher in amplitude in the equivalent harmonic range. We observe more distortion from the experiment for both geometries as the curves approach 4000 Hz. For spectral components below 900 Hz, Geometries 2 and 3 generate almost identical outputs. In particular, both underestimate the fundamental frequency of the $B_3^b$ by roughly 3 dB; but the second and third harmonics are roughly 8 dB and 3 dB too high, respectively.

Looking now at the $B_4^b$ note in the bottom plot of Figure \ref{fi:3d}, we see that the simulations run on Geometry 2 and Geometry 3 give similar solution curves. Again, Geometry 3 matches the experimental data almost perfectly for frequencies between 900 Hz - 2800 Hz. In addition, compared to Geometry 2, there is overall less variation between the experiment and the Geometry 3 solution. Regardless of the computational trumpet used, we observe more deviation as the numerical solutions approach frequencies near 4000 Hz.  Finally, for the lower spectrum, the fundamental frequency of the $B_4^b$ (which corresponds to the second harmonic of the $B_3^b$) is roughly 7 dB too high.

\begin{figure} [ht]
\centering
\includegraphics[width=7.5cm]{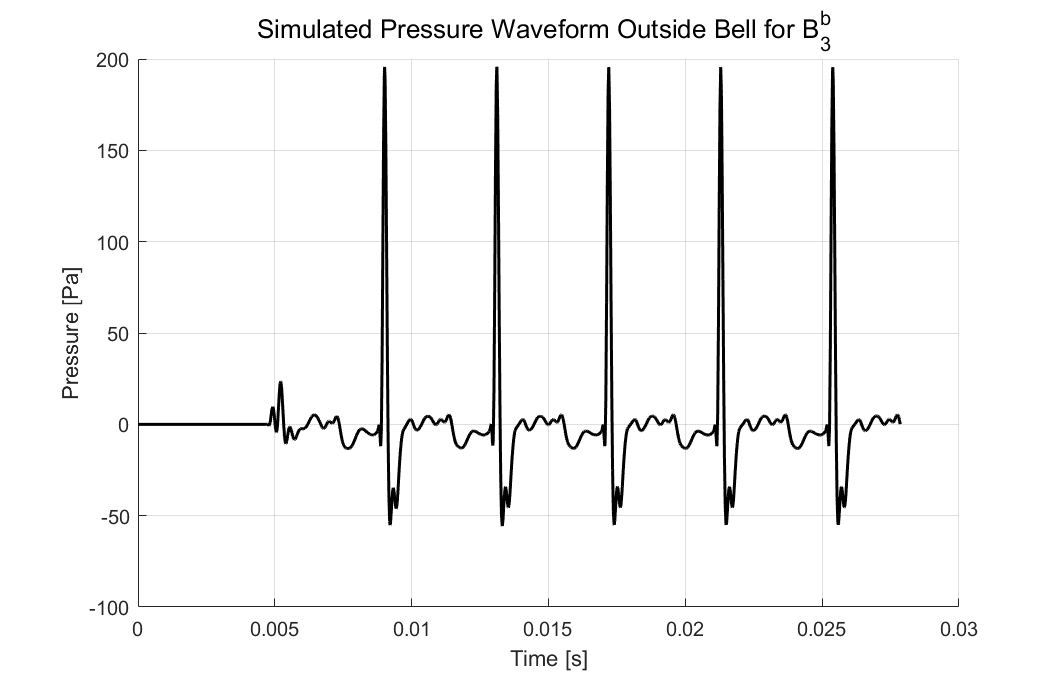} 
\caption[]{(Colour online)  Geometry 1 numerical results of $B_3^b$ outside the trumpet.}
\label{fi:Geo1B3}
\end{figure}

\begin{figure}[!ht]
\subfloat{\includegraphics[scale=0.28, width=7.5cm]{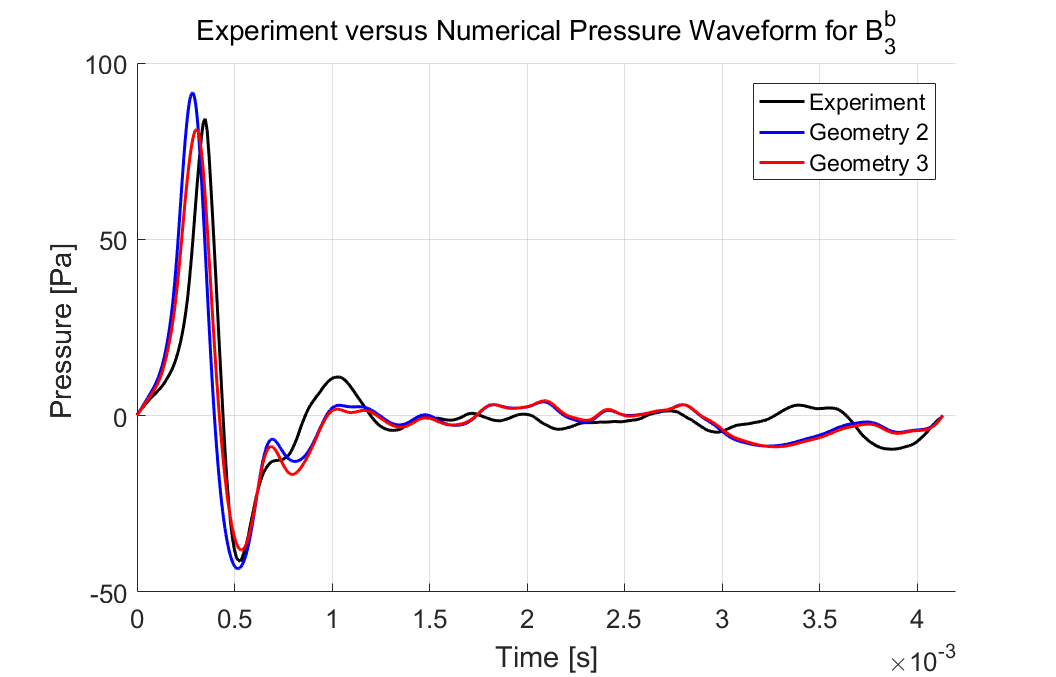}}
\qquad
\subfloat{\includegraphics[scale=0.28, width=7.5cm]{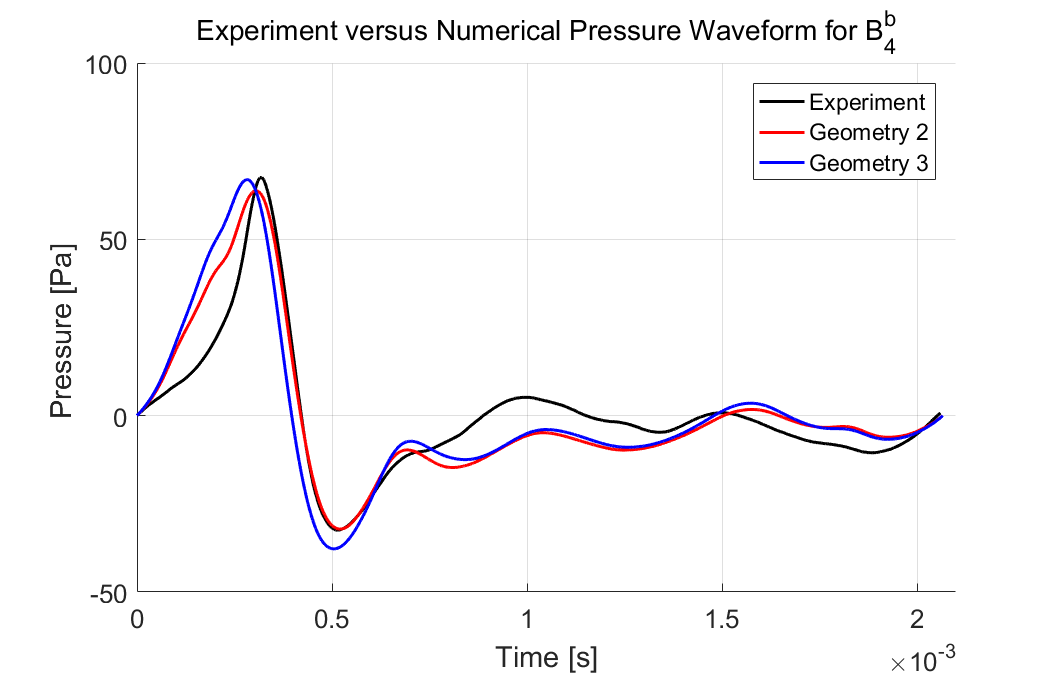}} 
\caption[]{(Colour online) Experimental and 3D numerical pressure waveform for $B_3^b$ (top) and $B_4^b$ (bottom).}
\label{fi:3dP}
\end{figure}

When we examined the numerical waveforms outside the bell, we found that the solutions reached steady state roughly halfway through the first period. We show an example of this in Figure \ref{fi:Geo1B3}, which depicts the Geometry 1 $B_3^b$ solution in the time-domain. By $ t = 0.00668$, the numerical result produces consistent periodic waveforms. Although only the first five periods of the $B_3^b$ curve is shown here, we have run simulations where more than a dozen periods were generated to ensure the solution was stable. Hence, the plots presented in Figures \ref{fi:3d} and \ref{fi:3dP} analyze the first period of the (results starting from the initial crest) since the spectra of the other periods were equivalent.  

In Figure \ref{fi:3dP}, we plot the measured pressure waveforms in the time-domain along with one period of the most accurate numerical results, i.e., the Geometry 2 and Geometry 3 solutions. From a compressible fluids simulation point of view, the numerical and experimental waveforms match rather well.

\begin{figure} [ht]
\centering
\includegraphics[width=7.5cm]{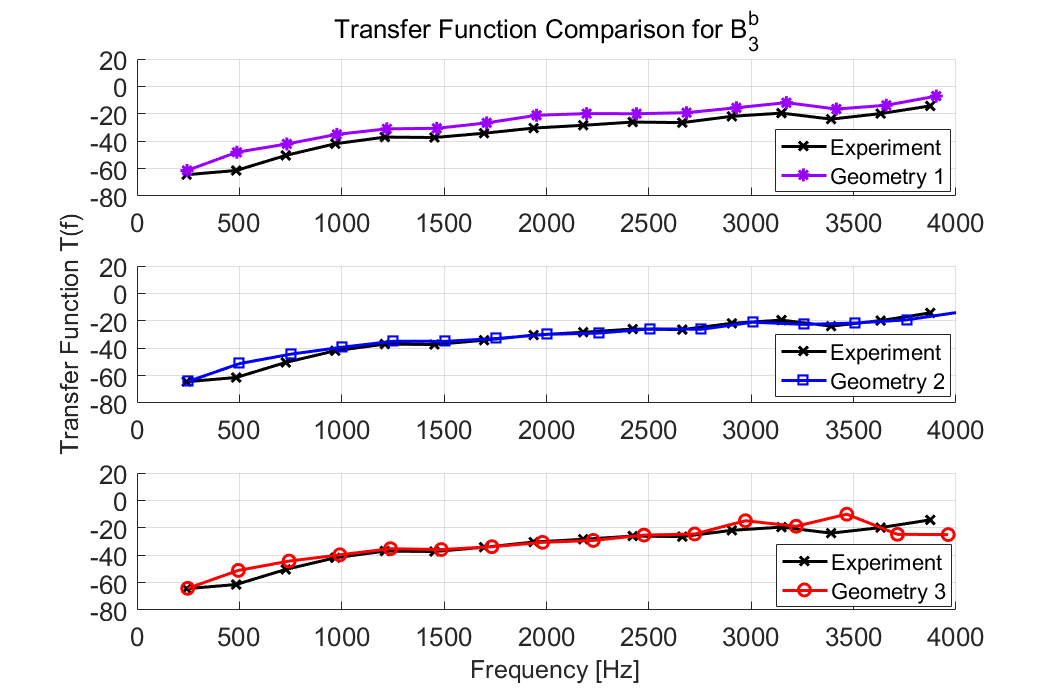} 
\caption[]{(Colour online) Transfer function of $B_3^b$ experimental data compared to simulation results, specifically the ratio of the pressure measured outside the bell over the pressure measured at the mouthpiece for Geometry 1 (top), Geometry 2 (middle) and Geometry 3 (bottom).}
\label{fi:TFB3}
\end{figure}

\begin{figure} [ht]
\centering
\includegraphics[width=7.5cm]{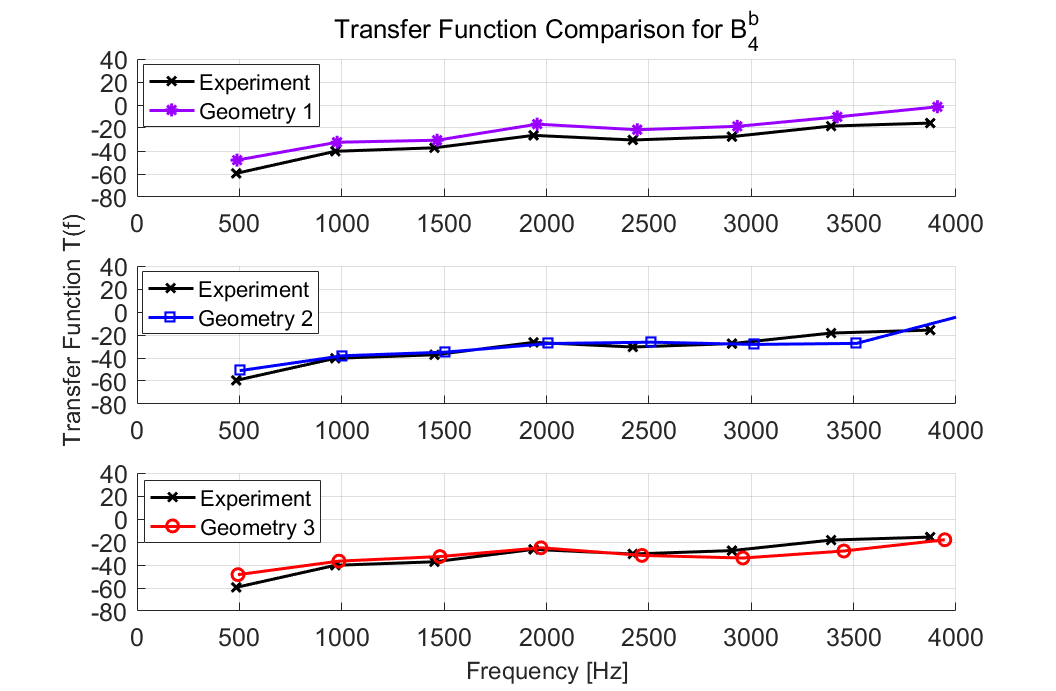} 
\caption[]{(Colour online) Transfer function of $B_4^b$ experimental data compared to simulation results, specifically the ratio of the pressure measured outside the bell over the pressure measured at the mouthpiece for Geometry 1 (top), Geometry 2 (middle) and Geometry 3 (bottom).}
\label{fi:TFB4}
\end{figure}

Finally, to examine the behaviour of the transmitted waves from the computational trumpet, the transfer function is calculated for all numerical results and compared to the experimental data. The transfer function for Geometry 1, Geometry 2 and Geometry 3 will be denoted by $T(f)_\text{Geo.1}$, $T(f)_\text{Geo.2}$ and $T(f)_\text{Geo.3}$, respectively. The corresponding functions are plotted in Figures \ref{fi:TFB3} and \ref{fi:TFB4} for $B_3^b$ and $B_4^b$, respectively. The top plots of Figures \ref{fi:TFB3} and \ref{fi:TFB4}, illustrate that $T(f)_\text{Geo.1}$ is similar to $T(f)_\text{Exp}$, but there is a significant difference in the amplitude value for all frequency components for both notes. Whereas $T(f)_\text{Geo. 2}$ and $T(f)_\text{Geo.3}$ shown in the middle and bottom plots, respectively, are more representative of $T(f)_\text{Exp}$. 

In summary, it is evident from our results that considering a 3D model of the trumpet where the initial bore geometry is considered greatly improves our numerical results, especially for the $B_3^b$. Our data shows spectral plots which agree with simulations over the lower harmonics of high amplitude, down to around 50 dB to 60 dB below the maximum  levels.  We interpret this to mean that our acoustic data is fairly good, but shows the effects of external noise for frequencies greater than about 4000 Hz. The measurement system is capable of over 90 dB of contrast, so it is not a limitation. Furthermore, our numerical results closely follow our data in this frequency range.

\vspace{-3mm}

\noindent \section{Conclusion}
\vspace{-1mm}

In conclusion, our 3D simulations of nonlinear wave propagation in a trumpet produce encouraging results. We found that the geometric modelling near the beginning of the trumpet is crucial to take into account, yet often is overlooked. We speculate the shape near the mouthpiece influences how the reflections of high pressure waves are modelled. The radiated frequencies of the Geometry 2 and 3 bells show evidence of this: although the computational trumpets have a comparable geometry and their spectra align well with the experimental data, there are distinguishable differences in their solution curves.

It is difficult to take precise measurements close to the mouthpiece; but not having an accurate geometry, even if the bore is uniform, produces numerical discrepancies. For instance, numerical experiments we did showed that enlarging the radius of the tube by a mere 0.45 mm shifts the frequency curve by almost 2 dB. If we compare the radius of the mouthpiece boundary for Geometry 1 to Geometries 2 and 3, it is 2.12 times larger than that of the other computational trumpets. Therefore, it seems reasonable that the corresponding numerical amplitude at the bell position is roughly 7 dB off $\left ( 20 \log_{10}{(2.12)} \approx 6.5 \text{ dB} \right )$.

\begin{figure}[!ht]
\subfloat{\includegraphics[scale=0.28, width=7.5cm]{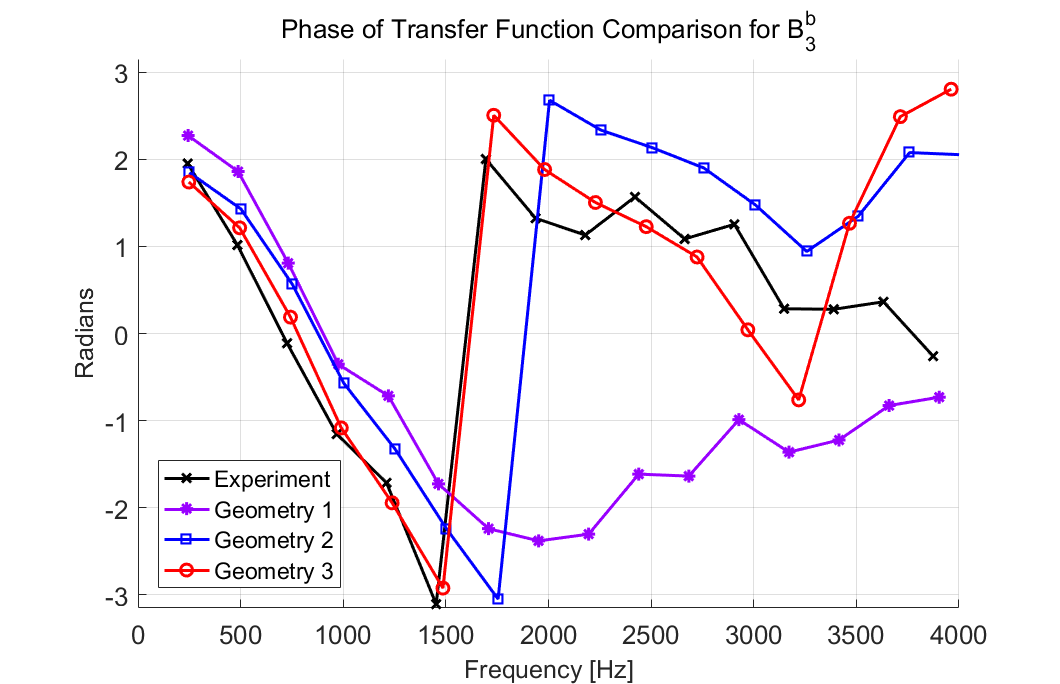}}
\qquad
\subfloat{\includegraphics[scale=0.28, width=7.5cm]{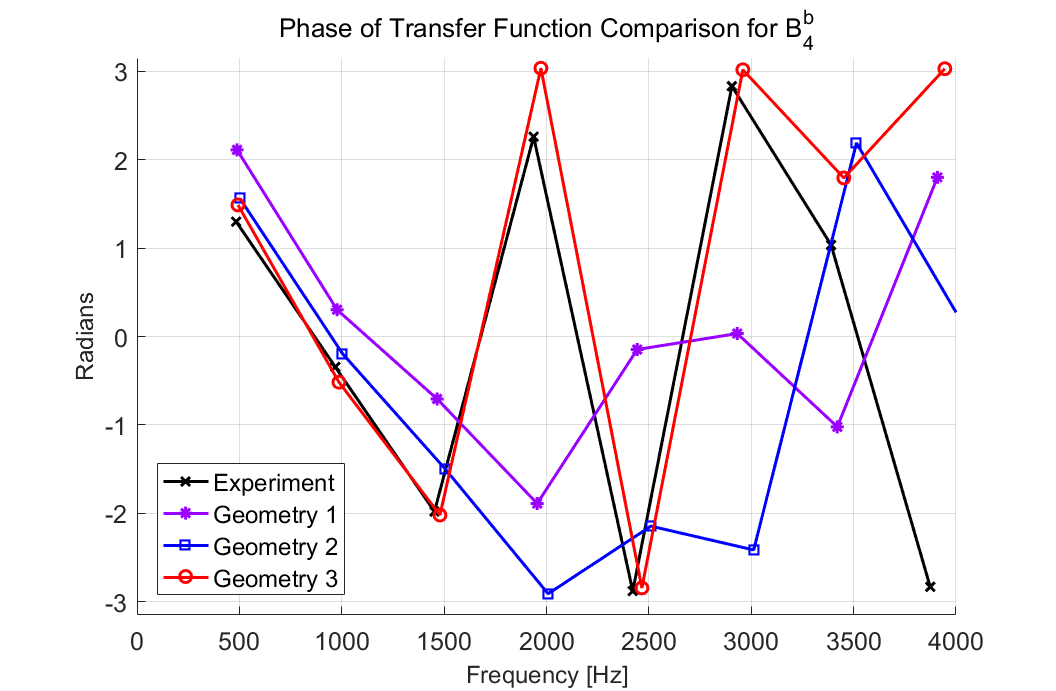}} 
\caption[]{(Colour online) Phase angle of transfer function for $B_3^b$ (top) and $B_4^b$ (bottom) of the experimental data compared to the simulation results for Geometry 1, Geometry 2 and Geometry 3. }
\label{fi:TFA}
\end{figure}

The lower harmonic components remain to be a challenge. For both notes the Geometry 2 and 3 simulations deviate the most for the frequency components near 400 Hz, i.e., the second harmonic and fundamental frequency of the $B_3^b$ and $B_4^b$, respectively. A large portion of the energy corresponding to frequencies around 400 Hz are reflected at the bell and travel toward the mouthpiece. So it seems there is an issue with how the reflections and corresponding phase angles are being modelled inside the bore.

In Figure \ref{fi:TFA}, we plot the phase angle of the transfer functions shown in Figures \ref{fi:TFB3} and \ref{fi:TFB4}. The graphs illustrate that the simulated phase angles match the measured for the lowest harmonic components. In particular, the phase angles corresponding to the $B_3^b$ and $B_4^b$ simulations run on Geometry 3 match those of the experiment for frequencies below 1800 Hz and 3000 Hz, respectively. However, for Geometries 1, 2 and 3, we do notice variations from the experiments. This implies the magnitude of crests and troughs may be altered, which in turn effects the shape of the standing waves and thus, whether wave steepening would occur. This ultimately could influence the brassiness of the instrument. 

Overall, we blame the difference is the lower spectra on the simplified model for velocity that was prescribed for the boundary condition. It is also possible that the lower frequencies are most influenced from energy losses. In \cite{Kausel15}, Kausel states that wall vibrations and full body motion of the bell influence the lower frequencies, especially between 350 Hz - 900 Hz.  Our future work will investigate how to improve this boundary condition to find a more accurate relationship between pressure and velocity.

\vspace{-2mm}
\subsection*{Acknowledgement}
\vspace{-1mm}

The authors thank Andrew Giuliani for his implementation of the DG method used for the simulations presented in this paper. This research was supported by the Natural Sciences and Engineering Research Council (NSERC) of Canada grants 341373 and 112608 and we acknowledge and thank NSERC for the financial support.

\vspace{-1mm}

\vspace{-1mm}
\bibliographystyle{plain}

\end{document}